\documentclass[prl,twocolumn,aps]{revtex4}
\usepackage{amsmath}
\usepackage{amssymb}
\usepackage{graphicx}

\begin{document}
\title{Gauge invariance of the local phase in the Aharonov-Bohm interference:
 quantum electrodynamic approach
}
\author{Kicheon Kang}
\email{kicheon.kang@gmail.com} 
\affiliation{Department of Physics, Chonnam National University, Gwangju 61186, 
 Republic of Korea}

\begin{abstract}
In the Aharonov-Bohm (AB) effect, interference fringes are observed 
for a charged particle in the absence of the local overlap with the external
electromagnetic field. This notion of the apparent ``nonlocality" of 
the interaction or the significant
role of the potential has recently been challenged and are under
debate. The quantum
electrodynamic approach provides a microscopic picture of the characteristics of
the interaction between a charge and an external field. 
We explicitly show the gauge invariance of the local
phase shift in the magnetic AB effect, which is in contrast to
the results obtained using the usual semiclassical vector potential. 
Our study can resolve the issue of the locality in the magnetic AB
effect. However, the problem is not solved in the same way
in the electric counterpart
wherein virtual scalar photons play an essential role.
\end{abstract}

\maketitle

%{\sf CHECK from HERE...} \\
{\em Introduction-}.
%\section{Introduction}
%
A charged particle moving under the influence of an external electromagnetic 
field exhibits a topological quantum interference, known as the 
Aharonov-Bohm~(AB) effect~\cite{ehrenberg49,aharonov59}. 
An intriguing aspect of the AB effect is that interference occurs
even when the particle does not locally overlap the field.
Thus, the AB interference is widely regarded 
as a pure topological effect which cannot be represented
by the local action of gauge-invariant quantities ({\em i.e.},
electromagnetic field). 
In our previous studies, we demonstrated that the notion of ``nonlocality" 
contradicts the prediction of local phase measurement 
in certain experimental arrangements~\cite{kang17,kim18}. 
The discrepancy could be resolved using a semiclassical approach based on 
the local
field interaction~\cite{kang13,kang15}, in which the external electromagnetic
field locally interacts with the field produced by a charged particle.

In recent years, it has been proposed that the local phase in the AB effect 
can be described by the quantum electrodynamic (QED) 
approach~\cite{marletto20,saldanha21}. In the QED picture, 
the interaction between a charge
and magnetic flux is mediated by the exchange of virtual photons,
represented by the gauge field $A^\mu$.
It has been claimed that the gauge-invariant local phase is 
derived from this approach. Although the QED picture of the
interaction between the two objects should ultimately be right,
the gauge invariance of the local phase remains unclear. This is because
the local phase has been derived by adopting specific choices of gauge 
in $A^\mu$ 
(e.g., Coulomb~\cite{marletto20} 
and Lorenz~\cite{saldanha21} gauges, respectively). 
The general gauge invariance should be proved to address this issue.

In this Letter, we demonstrate the gauge invariance of
the local phase in the magnetic AB effect using
the quantum electrodynamic approach. Moreover, 
the result is equivalent to that obtained using the semiclassical 
local field interaction approach~\cite{kang17}. 
The QED approach yields an unambiguous prediction of the 
local phase measurement
in a certain type of superconducting Andreev interferometer.
Our result is remarkable in that the 
semiclassical vector potential does not produce 
a well-defined gauge-invariant local phase.
%\\ ({\sf Some remarks on the meaning to be added ?})
%

{\em Quantum electrodynamic Hamiltonian of a charge and a fluxon-}.
The system considered comprises a charge $e$ and a 
"fluxon" $\Phi$ in two spatial dimension~(Fig.~1). 
This simplified configuration is sufficient to derive the essential physics
of the AB effect.  The interaction between the
two entities is indirect, {\em i.e.}, it is mediated by virtual photons 
(gauge field).  The Hamiltonian of the system can be written as
\begin{equation}
 H = \frac{1}{2m} (\mathbf{p} - \frac{e}{c}\mathbf{A})^2 
     + \frac{\mathbf{P}^2}{2M}
     - \frac{\Phi}{4\pi} \hat{z}\cdot\mathbf{B} 
     + \sum_{\mathbf{k},\lambda} \hbar\omega
       a_{\mathbf{k}\lambda}^\dagger a_{\mathbf{k}\lambda} \,,
\label{eq:H1}
\end{equation}
where the motion of the particles is confined in the $x$-$y$ plane.
The charge ($e$) with mass $m$ located at $\mathbf{x}_a$ interacts 
with the quantized radiation through the vector potential $\mathbf{A}$.
In contrast, the fluxon ($\Phi$) with mass $M$ at $\mathbf{x}_b$ 
interacts with the magnetic field $\mathbf{B}$ of the radiation.
The operator $a_{\mathbf{k}\lambda}^\dagger$ ($a_{\mathbf{k}\lambda}$)
represents the creation (annihilation) of a photon 
with wave number $\mathbf{k}$ and
polarization $\lambda$. %(limited to transverse modes). 
%$H_{\mathrm{em}}$ denotes the quantized radiation (photons).
Among the four possible modes of the polarization, only the two transverse
modes are real excitation of radiation (last term of Eq.~\eqref{eq:H1}).

The Hamiltonian can be rewritten as
\begin{subequations}
 \label{eq:H}
\begin{equation}
 H = H_0 + V_a + V_b ,
\end{equation}
where 
\begin{equation}
 H_0 = \frac{\mathbf{p}^2}{2m} + \frac{\mathbf{P}^2}{2M} 
     + \sum_{\mathbf{k},\lambda} \hbar\omega
       a_{\mathbf{k}\lambda}^\dagger a_{\mathbf{k}\lambda} 
\end{equation}
is the noninteracting part, and 
\begin{equation}
 V_a = -\frac{e}{mc} \mathbf{A}\cdot\mathbf{p}
\end{equation}
and
\begin{equation}
 V_b = - \frac{\Phi}{4\pi} \hat{z}\cdot\mathbf{B} 
  \label{eq:V_b}
\end{equation}  
\end{subequations}
represent the charge--vacuum and fluxon--vacuum interactions, respectively.
In the charge-potential interaction ($V_a$), we have omitted 
the term $(i\hbar/2mc) \nabla\cdot\mathbf{A} + (e^2/2mc^2)\mathbf{A}^2$ which
is independent of the charge variable and thus irrelevant to the present
study.

We can observe the asymmetry between $V_a$ and $V_b$ 
in the role of the gauge field
$\mathbf{A}$.
The charge interacts with the radiation $\mathbf{A}$, which has
some freedom of gauge selection. However, the interaction of the fluxon
with the radiation is described by the gauge-independent $\mathbf{B}$. 
The gauge independence of $V_b$ can be confirmed by considering the fluxon 
as a current loop (with current $I$) 
interacting with $\mathbf{A}$ as
\begin{equation}
 V_b = - \frac{I}{c} \oint \mathbf{A}\cdot d\mathbf{l} \,,
\end{equation} 
where the integration is made along the loop. Applying the Stokes' theorem,
we recover Eq.\eqref{eq:V_b}, which is independent of gauge choice in 
$\mathbf{A}$.

The vector potential $\mathbf{A}$ corresponds to the spatial part
of the four-potential $A^\mu$. The case of arbitrary
gauge will be treated later. First, we begin with the Coulomb gauge 
($\nabla\cdot\mathbf{A}=0$), where
the two transverse modes (represented by $\lambda$) of $A^\mu$ 
are present:
\begin{equation}
 \mathbf{A}(\mathbf{x},t) = 
  \sum_{\mathbf{k},\lambda} \alpha_\mathbf{k} 
  \left[ 
    u_\mathbf{k}(\mathbf{x}) a_{\mathbf{k}\lambda} e^{-i\omega t}
   +u_\mathbf{k}^*(\mathbf{x}) a_{\mathbf{k}\lambda}^\dagger e^{i\omega t}
  \right] \hat{e}_\lambda \,,
\label{eq:A}
\end{equation}
expanded by the plane wave modes
$u_\mathbf{k}(\mathbf{x}) =  e^{i\mathbf{k}\cdot\mathbf{x}}/\sqrt{V}$
with normalization coefficient 
$\alpha_\mathbf{k} = \sqrt{2\pi\hbar c^2/\omega}$ and the corresponding
angular velocity $\omega = ck$. 
The polarization $\hat{e}_\lambda$ is limited to the transverse modes 
by the constraint $\mathbf{k}\cdot\hat{e}_\lambda = 0$  in
the Coulomb gauge. 
%That is, the longitudinal and the scalar components 
%of $A^\mu$ are not present. 
The magnetic field of the radiation, given by 
$\mathbf{B} = \nabla\times\mathbf{A}$, is gauge-independent.
\begin{equation}
 \mathbf{B}(\mathbf{x},t) = i
  \sum_{\mathbf{k},\lambda} k\alpha_\mathbf{k} 
  \left[ 
    u_\mathbf{k}(\mathbf{x}) a_{\mathbf{k}\lambda} e^{-i\omega t}
   -u_\mathbf{k}^*(\mathbf{x}) a_{\mathbf{k}\lambda}^\dagger e^{i\omega t}
  \right] \hat{n}_\lambda \,,
\end{equation}
where $\hat{n}_\lambda = \hat{\mathbf{k}}\times \hat{e}_\lambda$.

{\em Canonical transformation and effective interaction-}. \\
We adopt the canonical transformation technique to derive the charge--fluxon 
interaction mediated by the exchange of virtual photons.
The Hamiltonian in \eqref{eq:H} is transformed to 
$\tilde{H} = e^{-S} H e^{S}$, 
where the first-order interaction part is eliminated by an appropriate 
choice of $S$. We obtain
\begin{subequations}
\label{eq:H_eff}
\begin{eqnarray}
 \tilde{H} &=& H_0 + H_2 \,, \\
  H_2 &=& \sum_\gamma |0\rangle 
   \frac{ \langle0|V_a+V_b|\gamma\rangle\langle\gamma|V_a+V_b|0\rangle }{
        -\hbar\omega}  \langle0| \,,
\end{eqnarray} 
where $|0\rangle$ denotes the radiation vacuum and $|\gamma\rangle = 
a_{\mathbf{k}\lambda}^\dagger|0\rangle$.
The self-interaction terms containing
$\langle0|V_a|\gamma\rangle\langle\gamma|V_a|0\rangle$ 
or $\langle0|V_b|\gamma\rangle\langle\gamma|V_b|0\rangle$ 
are irrelevant and thus discarded. The second-order interaction between
the two particles is given by 
\begin{equation}
 \tilde{H}_2 = \sum_\gamma |0\rangle 
   \frac{ \langle0|V_a|\gamma\rangle\langle\gamma|V_b|0\rangle 
         + \mathrm{h.c.} }{ -\hbar\omega}  \langle0| \,.
\label{eq:tilde_H2}
\end{equation}
\end{subequations}

The evaluation of the matrix elements in Eq.~\eqref{eq:tilde_H2} yields
\begin{subequations}
\begin{eqnarray}
 \langle0|V_a|\gamma\rangle &=& -\frac{e}{mc}\alpha_\mathbf{k} 
   u_\mathbf{k}(\mathbf{x}_a) (\hat{e}_\lambda\cdot\mathbf{p}) e^{-i\omega t} 
    \,, \\
 \langle\gamma|V_b|0\rangle &=& i\frac{\Phi}{4\pi} k\alpha_\mathbf{k}
   u_\mathbf{k}^*(\mathbf{x}_b) (\hat{z}\cdot\hat{n}_\lambda) e^{i\omega t} \,,
\label{eq:Vb}
\end{eqnarray}
\end{subequations}
and  we obtain
\begin{subequations}
\begin{equation}
 \tilde{H}_2 = \frac{ie\Phi}{2mc} 
   F(\mathbf{x}_a-\mathbf{x}_b) + \mathrm{h.c.} \,,
\end{equation}
where $F$ is a function of the relative position of the two objects
given by
\begin{equation}
 F(\mathbf{x}) = \frac{1}{4\pi^2} \int d^2\mathbf{k} \frac{1}{k}
  e^{i\mathbf{k}\cdot\mathbf{x}}
  \sum_\lambda (\hat{e}_\lambda\cdot\mathbf{p})(\hat{z}\cdot\hat{n}_\lambda) 
 \,.
\end{equation}
\end{subequations}
Applying the condition $\hat{e}_\lambda\cdot\mathbf{k}=0$ in the Coulomb gauge,
we find 
\begin{displaymath}
 (\hat{e}_\lambda\cdot\mathbf{p})(\hat{z}\cdot\hat{n}_\lambda) 
  = \hat{\phi}\cdot\mathbf{p} \,,
\end{displaymath}
where $\hat{\phi}$ is the angular unit vector in the space of $\mathbf{k}$.
Finally we can derive $\tilde{H}_2$ as
\begin{subequations}
\begin{equation}
 \tilde{H}_2 = -\frac{e}{mc} \mathbf{p}\cdot\mathbf{a} \,,
\label{eq:tildeH2}
\end{equation}
where the ``effective vector potential" $\mathbf{a}$ is given by
\begin{equation}
 \mathbf{a}(\mathbf{x}) = \frac{\Phi}{2\pi|\mathbf{x}|} \hat{\theta} 
\label{eq:a}
\end{equation} 
\end{subequations}
($\hat{\theta}$ denotes the azimuthal unit vector of $\mathbf{x}\equiv
\mathbf{x}_a-\mathbf{x}_b$).
This yields the expression of the effective Hamiltonian of 
Eq.~\eqref{eq:H_eff}
\begin{equation}
 \tilde{H} = \frac{1}{2m} \left( \mathbf{p}-\frac{e}{c}\mathbf{a} \right)^2
 \,,
\end{equation}
where terms independent of the interaction between the the charge and the
fluxon are omitted.

%{\sf The effective vector potential was obtained by others...}
The expression of the ``effective vector potential" in Eq.~\eqref{eq:a}
was previously obtained by
adopting the QED approach~\cite{santos99,lee04}, 
and it has recently been
addressed in the context of the locality of 
the interaction~\cite{marletto20,saldanha21}.
Marletto and Vedral~\cite{marletto20} and Saldanha~\cite{saldanha21} 
claimed that they had shown the locality of the interaction 
with particular selection of gauges 
(Coulomb and the Lorenz gauges, respectively).
However, this claim is incomplete without an explicit derivation of the 
gauge invariance. % of $\mathbf{a}$ in Eq.~\eqref{eq:a}. 
This is clear if we compare it to the
semiclassical approach where 
the vector potential $\mathbf{A}_\Phi$ produced
by a magnetic flux possesses some degree of choosing the gauge 
with the constraint $\oint\mathbf{A}_\Phi\cdot d\mathbf{x} = \Phi$. 
The gauge invariance of $\mathbf{a}$ leads to a notable consequence in
real experiments.
Without the gauge invariant $\mathbf{a}$ 
we cannot predict the value of the local phase shift in a certain experimental 
arrangement~\cite{kang17}, namely, the AB effect without an AB
loop. %(See also Ref.~\onlinecite{marletto20,saldanha21}).

{\em Gauge invariance of the interaction Hamiltonian and the local phase 
measurement-}.
Consider a gauge transformation of the four-potential
$A^\mu\rightarrow A'^\mu = A^\mu + \partial^\mu\Lambda$ with an arbitrary
single-valued scalar function $\Lambda = \Lambda(\mathbf{x},t)$.
The vector potential $\mathbf{A}$ in Eq.~\eqref{eq:A} is then transformed
to 
\begin{subequations}
\label{eq:A'}
\begin{equation}
 \mathbf{A}' = \mathbf{A} + \nabla\Lambda \,,
\end{equation}
where $\Lambda$ can be expanded as
\begin{equation}
 \Lambda(\mathbf{x},t) = \sum_{\mathbf{k},\lambda} 
  \left[
    \gamma_\mathbf{k} u_\mathbf{k}(\mathbf{x}) 
     a_{\mathbf{k}\lambda} e^{-i\omega t}
   + \gamma_\mathbf{k}^* u_\mathbf{k}^*(\mathbf{x}) 
     a_{\mathbf{k}\lambda}^\dagger e^{i\omega t}
  \right] \,,
\end{equation}
\end{subequations}
with an arbitrary $\mathbf{k}$-dependent coefficient $\gamma_\mathbf{k}$.

The additional contribution 
to $\tilde{H}_2$ of Eq.~\eqref{eq:tilde_H2} may be produced by
$\partial^\mu\Lambda$ in the gauge transform.
First, scalar potential $A^0$ is nonzero unlike in the Coulomb gauge.
However, it does not couple to the motion of charge $\mathbf{p}$
and is irrelevant to the interaction between $q$ and $\Phi$.
Second, the spatial part $\nabla\Lambda$ generates an additional term 
in $\tilde{H}_2$, given by 
\begin{subequations}
\begin{equation}
 \delta\tilde{H}_2 = \sum_\gamma |0\rangle
   \frac{ \langle0|\delta V_a|\gamma\rangle\langle\gamma|V_b|0\rangle
         + \mathrm{h.c.} }{ E_0-E_\gamma}  \langle0| \,,
\end{equation}
where 
\begin{equation}
 \delta V_a = -\frac{e}{mc} \nabla\Lambda\cdot \mathbf{p} \,.
\end{equation}
$\nabla\Lambda$ consists of the longitudinal modes 
(parallel to $\mathbf{k}$), and we obtain 
\begin{equation}
 \langle0|\delta V_a|\gamma\rangle = -\frac{ie}{mc} k\gamma_\mathbf{k} 
    u_\mathbf{k}(\mathbf{x}_a) (\hat{\mathbf{k}}\cdot\mathbf{p}) 
    e^{-i\omega t}.
\end{equation}
\end{subequations}
However, this term does not couple to $\langle\gamma|V_b|0\rangle$, because
the latter contains only transverse components (see Eq.~\eqref{eq:Vb}). 
Therefore we conclude that $\delta\tilde{H}_2 = 0$. 
{\em The second order interaction
$\tilde{H}_2$ is gauge invariant}.

This result of deriving gauge invariance in $\tilde{H}_2$ is
remarkable.  First, the gauge-invariant effective vector potential
($\mathbf{a}$ in Eq.~\eqref{eq:a}) is in sharp contrast 
with the semiclassical counterpart. 
In the semiclassical approach, a charged
particle under an external magnetic field is described by the Hamiltonian
\begin{equation}
 H = \frac{1}{2m} \left( \mathbf{p}-\frac{e}{c}\mathbf{A}_\Phi \right)^2
 \,,
\label{eq:H_APhi}
\end{equation}
where the vector potential $\mathbf{A}_\Phi$, generated by $\Phi$, 
can be transformed to another function 
$\mathbf{A}_\Phi \rightarrow \mathbf{A}_\Phi + \nabla\chi(\mathbf{x})$.
The constraint in $\mathbf{A}_\Phi$ is not local: 
$\oint \mathbf{A}_\Phi \cdot d\mathbf{r} = \Phi$.

Second, the gauge-invariant $\tilde{H}$ provides the unambiguous prediction
of the local phase that can be measured in a certain type of experimental
arrangement~\cite{kang17}. 
The Hamiltonian~\eqref{eq:H_APhi} with semiclassical 
$\mathbf{A}_\Phi$ fails to make this prediction.
Here, we briefly review the essential feature
of the local phase measurement experiment 
(For the details, see Ref.~\onlinecite{kang17}). Its schematic setup 
consists of two independent superconducting leads ($S_1$ and $S_2$),
which are tunnel-coupled to a common normal electrode~($N$) (Fig.~2). 
The two superconducting electrodes are biased with identical voltages below the
superconducting gap. The electrical current flows to the normal metallic output
via Andreev reflection~(AR)~\cite{andreev64},
in which a Cooper pair is converted to two normal electrons in $N$.
Interference is caused by the indistinguishability of the
two different AR processes ($S_1$ to $N$ or $S_2$ to $N$), and
the phase shift produced by the external flux is 
\begin{equation}
 \phi_\mathrm{loc} = \frac{e^*}{\hbar c} \int_C \mathbf{a}\cdot d\mathbf{x} 
   = \frac{e^*\Phi}{2\pi\hbar c} \Delta\theta \,,
 \label{eq:phi_loc}
\end{equation}
where the integration is considered along path $C$, as shown in 
Fig.~2. The effective charge 
$e^*=2e$ corresponds to the charge of a Cooper pair, and $\Delta\theta$ is 
the angle formed in the geometry of the system. 
The superconducting state is associated with gauge symmetry
breaking, and the Cooper pairs satisfy the bosonic statistics. Therefore,
the charge conservation and the Fermionic superselection rule do not
prevent the interference in this setup 
(contrary to the argument in Ref.~\onlinecite{marletto20}).

Third, this value of the local phase (Eq.~\eqref{eq:phi_loc}) is equivalent
to the result predicted by the semiclassical local field interaction 
(LFI) approach~\cite{kang17}. As mentioned above, this cannot be achieved
based on the semiclassical vector potential~(Eq.~\eqref{eq:H_APhi}).
With $\mathbf{A}_\Phi$, the local phase is given by
$ \phi_\mathrm{loc} = (e^*/\hbar c) \int_C \mathbf{A}_\Phi\cdot d\mathbf{x}$,
and it is not invariant under the gauge transformation 
$\mathbf{A}_\Phi \rightarrow \mathbf{A}_\Phi + \nabla\chi$.
This implies that the potential-based semiclassical approach 
fails to predict the value of $\phi_\mathrm{loc}$.

We need to address about how the ambiguity of the charge-flux interaction
could be eliminated in the QED approach 
(represented by the effective vector potential 
$\mathbf{a}$ in Eq.~\eqref{eq:a}).
The key point is that the semiclassical Hamiltonian of 
Eq.~\eqref{eq:H_APhi} does not include the information regarding the local
configuration of the system. Owing to the freedom of selecting a gauge
in $\mathbf{A}_\Phi$, the same configuration of the flux can be
described by a different function for $\mathbf{A}_\Phi$; 
furthermore, different
locations of the flux may be expressed by the same $\mathbf{A}_\Phi$.
In contrast, transformation of the gauge field 
(Eq.~\eqref{eq:A'}) in the QED approach does not change 
the interaction Hamiltonian (Eq.~\eqref{eq:tildeH2}).
Its difference from the semiclassical vector potential is twofold.
First, the local configuration of the charge and flux is well specified
in the QED approach. Second,
$\nabla\Lambda$ in the transformation of $\mathbf{A}$ (Eq.~\eqref{eq:A'})
contains only the longitudinal modes in the
gauge field. The longitudinal modes do not contribute to the
interaction because the magnetic flux is coupled to the transverse modes.

It was recently proposed that the local phase $\phi_\mathrm{loc}$ can be 
predicted in the QED approach based on the particular choices of 
gauge~\cite{marletto20,saldanha21}.
As described above, this claim should be supported 
by an explicit demonstration of the gauge invariance of $\mathbf{a}$. 
This is evident if we compare it to the semiclassical vector potential 
$\mathbf{A}_\Phi$ in Eq.~\eqref{eq:H_APhi}, where such a gauge invariance
is not satisfied. We have
explicitly shown the gauge invariance of the effective vector potential
and confirmed that the QED approach eliminates the ambiguity in
the interaction strength of the charge and flux.

{\em Scalar photons and the scalar Aharonov--Bohm effect-}.
The gauge invariance of the effective vector potential $\mathbf{a}$ raises
a question whether this invariance is universal for any type of
the electromagnetic interaction mediated by virtual photons. In the following,
we show that this is not the case if the scalar photons come into play.
Consider a stationary charge ($e$) under external charge distribution ($\rho$)
in the ideal force-free condition, as shown in Fig.~3. 
This condition can be achieved in real 
experiments~\cite{aharonov59,kim18}, though it has never been realized.
The generation of external potential with a vanishing electric field
requires an appropriate distribution
of charge density $\rho(\mathbf{x}')$. In practice, this condition is
achieved in the Faraday cage. The interaction between $e$ and
$\rho$ is mediated by virtual photons, and the system is described by
the Hamiltonian
\begin{subequations}
\begin{equation}
 H = H_0 + eV(\mathbf{x},t) + \int \rho(\mathbf{x}') V(\mathbf{x}',t) \,,
\end{equation} 
where $H_0$ represents the noninteracting part of $e$, $\rho$, 
and the electromagnetic
vacuum, respectively. The charges interact with the scalar potential $V$
(time component of $A_\mu$) of the radiation
\begin{equation}
  V(\mathbf{x},t) = A_0(\mathbf{x},t) =
  \sum_\mathbf{k} \alpha_\mathbf{k}
  \left[
    u_\mathbf{k}(\mathbf{x}) a_{\mathbf{k}0} e^{-i\omega t}
   +u_\mathbf{k}^*(\mathbf{x}) a_{\mathbf{k}0}^\dagger e^{i\omega t}
  \right] \,,
\label{eq:V}
\end{equation}
\end{subequations}
where $a_{\mathbf{k}0}$ ($a_{\mathbf{k}0}^\dagger$) annihilates (creates)
a photon in the scalar mode.
Adopting the same canonical transformation technique used above (in obtaining 
Eq.~\eqref{eq:H_eff}), we can derive the Coulomb interaction mediated by the
virtual scalar photons~(see e.g., Ref.~\onlinecite{zee10}), as
\begin{equation}
 \tilde{H} = H_0 + e\int \frac{\rho(\mathbf{x}')}{|\mathbf{x}-\mathbf{x}'|}
   d^3\mathbf{x}' \,.
\label{eq:H_Coulomb}
\end{equation}
This expression of the Coulomb interaction is derived from the exchange 
of virtual photons in the scalar mode.
It is obtained by imposing the Lorenz gauge condition 
$\partial_\mu A^\mu = 0$ 
and is not fully gauge-invariant. For instance, if we choose 
the Coulomb gauge, the scalar and
longitudinal modes are absent. In this case, $V=0$ in Eq.~\eqref{eq:V}
and the Coulomb interaction of Eq.~\eqref{eq:H_Coulomb} cannot be derived.

This implies that the question regarding the reality of $A^\mu$
is more subtle in the quantum electrodynamics involving scalar modes.
In any case, there are two notable points:
The scalar photons (i) can never be observed
but (ii) are indispensable as the mediator of the Coulomb interaction
between two separate charges.

{\em Conclusion-}.
In the quantum electrodynamic approach to the AB effect, the interaction
between a charge and a magnetic flux is mediated by the exchange of virtual
photons. We have shown the gauge invariance of the charge-flux interaction,
which leads to an unambiguous prediction of the local phase shift.
%which is unavailable in the usual semiclassical vector potential approach. 
We have shown 
that the problem is more subtle in the electric AB effect. 
In contrast to the case of the magnetic AB, the scalar component of the
gauge field plays an essential role and cannot be gauged away 
in the electric AB effect. This may indicate a significant role of the gauge
field beyond mathematical construction.

%\begin{thebibliography}{99}
 \bibliography{references}
%\end{thebibliography}

\newpage
%\begin{figure}[l]
\begin{figure}
\centering
\includegraphics[width=4cm]{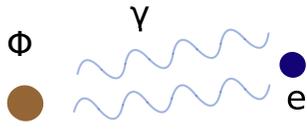} 
\caption{Charge ($e$) and fluxon ($\Phi$) in the vacuum. 
The mutual interaction between
the two objects is mediated by the vacuum radiation field (represented by
$\gamma$).}
\end{figure}
\begin{figure}
\centering
\includegraphics[width=7cm]{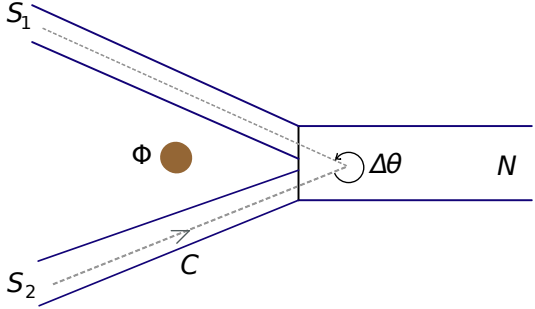} 
\caption{Schematic setup of the Andreev interferometer comprising
two superconducting electrodes ($S_1$ and $S_2$) in contact with
a normal metal ($N$). }
\end{figure}
%
%\begin{figure}[l]
\begin{figure}
\centering
\includegraphics[width=5cm]{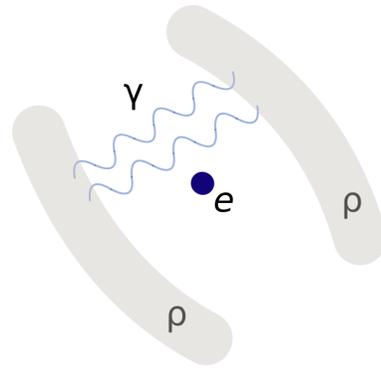} 
\caption{A charge ($e$) and the external charge distribution (represented
by the charge density $\rho$). The scalar component of gauge field mediates the
Coulomb interaction between $e$ and $\rho$.}
\end{figure}
\end{document}